\documentclass[11pt]{article}

\usepackage{latexsym,amsfonts,amsmath,amssymb,wasysym,enumerate,epsfig}
\usepackage{theorem}

\numberwithin{equation}{section}

\newcommand{\beq}{\begin{equation}}
\newcommand{\eeq}{\end{equation}}
\newcommand{\bea}{\begin{eqnarray}}
\newcommand{\eea}{\end{eqnarray}}

\newcommand{\RR}{{\mathbb R}}
\newcommand{\CC}{{\mathbb C}}
\newcommand{\eps}{\epsilon}

\begin{document}

\renewcommand{\thefootnote}{\arabic{footnote}}
\setcounter{footnote}{0}
\newpage
\setcounter{page}{0}

\pagestyle{empty}

\begin{center}

\begin{minipage}{13cm}
{\Huge \textsf{Quantum spins on star graphs  
\vspace{2mm}\\
and the Kondo model}}\\

\vfill

{\large \textbf{N. Cramp{\'e}}$^{a,b}$ and \textbf{A. Trombettoni}$^{c,d}$ }\\

\vfill

\emph{$^a$ CNRS, Laboratoire Charles Coulomb UMR 5221, Place Eug{\`e}ne Bataillon - CC070, F-34095 Montpellier, France}

\emph{$^b$ Universit\'e Montpellier II, Laboratoire Charles Coulomb UMR 5221, F-34095 Montpellier, France}

\emph{$^c$ CNR-IOM DEMOCRITOS Simulation Center and SISSA, 
Via Bonomea 265, I-34136 Trieste, Italy} 

\emph{$^d$ INFN, Sezione di Trieste, I-34127 Trieste, Italy}
\vspace{5mm}\\
E-mail: nicolas.crampe@univ-montp2.fr, andreatr@sissa.it\\
\vspace{2cm}
\begin{abstract}
We study the XX model for quantum spins on the star graph with three legs 
(i.e., on a $Y$-junction).
By performing a Jordan-Wigner 
transformation supplemented by the introduction of an auxiliary space  
we find a Kondo Hamiltonian of fermions, in the spin $1$ representation of $su(2)$, 
locally coupled with a magnetic impurity. In the continuum limit our model 
is shown to be equivalent to the $4$-channel Kondo model coupling 
spin-$1/2$ fermions with a spin-$1/2$ impurity and exhibiting 
a non-Fermi liquid behavior. 
We also show that it is possible to find a XY model 
such that - after the Jordan-Wigner transformation - one obtains a 
quadratic fermionic Hamiltonian directly diagonalizable. 
\end{abstract}
\end{minipage}

\vfill
\end{center}
\centerline{Keywords: Quantum spin model;~Kondo model;~Jordan-Wigner transformation;~Star graph}
\vfill
\centerline{PACS: 02.03.Ik;~75.10.Dg;~75.30.Mb}
\vfill
\newpage

\section{Introduction}

Spatial inhomogeneities and their role in the emergence of 
coherent behaviors at mesoscopic length scales are the subject 
of a continuing interest. In general, 
spatial inhomogeneities may be 
random, due to the presence of disorder or noise, 
as well as non-random, as a result of an external control on the 
geometry of the system: in a broad sense, their formation 
can be dynamically generated 
or induced through a suitable engineering of the system. 
As a consequence the effects of spatial inhomogeneities have been investigated 
in a variety of systems, ranging from pattern formation in 
systems with competing interactions \cite{Cha11} to 
Josephson networks with non-random, yet non-translationally invariant 
architecture \cite{Bur00,Sod06}.

A paradigmatic system in which the effects of spatial inhomogeneities 
can be studied is provided by spin models: on the one hand, not only do spin Hamiltonians 
directly describe many phenomena of magnetic systems \cite{Mat06}, 
including the effects of frustration \cite{Lac11}, 
but they are also routinely used to model 
physical properties of several condensed matter systems. 
On the other hand, spatial inhomogeneities can be straightforwardly 
included in spin models, to explore the consequences of the breaking 
of the translational invariance and the local properties 
on the length scales of the inhomogeneities \cite{Igl}.
 
As an example of the application of the study of spatial 
inhomogeneities in spin systems, we mention the spin chain Kondo effect. 
The standard Kondo effect arises from the interactions 
between magnetic impurities and
the electrons in a metal and it is 
characterized by a net increase at low temperature
of the resistance \cite{Kon,Hew93,Kou01}. 
The Kondo effect has been initially observed for metals, like copper,
in which magnetic atoms, like cobalt, are added: 
however, interest in the Kondo physics
persisted also because it can be studied with quantum dots
\cite{Ali96,Kou98}. 
The universal low-energy/long-distance
physics of the Kondo model can be simulated and studied by a magnetic impurity
coupled to a gapless antiferromagnetic one-dimensional chain having 
nearest- and next-nearest- neighbour couplings $J_1$-$J_2$ \cite{Laf08}, 
with the the correct scaling behavior of the single channel Kondo problem 
being exactly reproduced by this spin model only when $J_2$ equals a critical 
value \cite{Laf08}. The spin model reproducing the low-energy behaviour 
of the Kondo problem is defined on the half line, since 
the radial coordinate of the fermionic model as well 
varies in the half line: the 
rationale is that the free electron Kondo problem
may be described by a one-dimensional model since
only the $s$-wave part of the electronic wavefunction is affected by the
Kondo coupling \cite{Aff08}. 
Another example in which the scaling behavior of a fermionic 
Kondo model may be well reproduced by a pertinently chosen spin model 
is discussed in \cite{Bay12}. Using the spin chain 
version of the Kondo problem, 
a characterization of the Kondo regime using negativity was recently 
presented \cite{Bay12,Bay10} and it was shown that long-range entanglement 
mediated by the Kondo cloud can be induced by a quantum quench \cite{Sod10}. 
It stands as a open and interesting line of research to introduce and study 
spin systems, eventually with suitably tailored spatial inhomogeneities, 
reproducing the scaling behaviour of more general systems of fermions 
coupled to magnetic impurities, as the general multichannel Kondo effect.

Another reason of interest for introducing spatial inhomogeneities in 
spin models defined on networks is given by the study of the effects 
of the topology 
of the graph on the properties of the system 
and of the breaking of integrability. As a main example, 
consider a quantum (classical) spin model which is 
integrable in one (two) dimensions. Techniques have been developed to deal 
with open boundary conditions \cite{CardyBook}, as for 
free boundaries described by algebraic curves 
\cite{Cardy}. However, if some vertices of the graph 
on which the spins are located have a number of nearest neighbours  
larger than all others, then integrability is in general broken. 
One can see this by considering a one-dimensional quantum 
model which can be solved by a Jordan-Wigner (JW) 
transformation \cite{JW}: intersecting the chain at one site
with a finite or infinite 
number of other chains the usual JW transformation on the spin 
variables will produce a fermionic model which is in general 
neither quadratic nor 
local. We recall that the two-dimensional classical Ising model at finite 
temperature can be solved by writing its partition functions in terms of a 
suitable quantum spin model on the chain which is solved by JW 
transformation \cite{LSM2,Muss}. Therefore, finding an effective way 
of performing a JW transformation in non trivial graphs 
amounts to the possibility 
of studying and possibly solving the Ising model in some non trivial 
(non two-dimensional) lattices \cite{DM}.

In this paper we study the XX model on a star graph obtained by merging 
three chains: the standard JW transformation 
cannot work for a star graph since there is no natural order on it 
(of course, this problem would generically appear for any graph, except for the 
circle and the segment where it works). 
We rather found convenient to supplement the application of the standard 
JW transformation with the introduction of an auxiliary space: the 
procedure of adding auxiliary sites to perform a JW transformation 
has been recently used to study higher-dimensional systems in \cite{Ga,Ci}. 
In our case, it is the use of this auxiliary space which allows us to get a 
Hamiltonian that is both quadratic and local in the JW fermions. 
Using such exact mapping, we show that the XX model on a star graph 
is equivalent to a generalized Kondo model, where the JW fermions 
enter locally and quadratically, and are coupled to a magnetic impurity. 

There are several reasons for our choice of the XX model on a star graph. 
On the one hand, we study the XX model since we are motivated by the need 
of emphasize the main point 
of our construction in the simplest case: for the XX model 
in a chain, the JW transformation gives rise to free fermions 
(our construction can be extended to other spin models solvable by JW 
transformations). 

On the other hand, we decide to restrict ourselves to the study of a star graph with three legs 
for a twofold reason: first, it is the simplest graph which 
can be constructed by merging a finite number of chains and having a finite 
number of vertices (three in our construction, see Figure \ref{fig:star}) 
with coordination number $z=3$ different and larger than the others 
(having $z=2$, with the sites at the boundaries of the chain having 
$z=1$). Second, the star graph (alias, the 
$Y$-junction) 
has been deeply studied in different contexts from different point of views: 
for three Tomonaga-Luttinger liquids (TLL) crossing at a point 
new attractive fixed points emerge \cite{Lal,COA,Giul}. Regular networks of 
TLL, with each node described by a unitary scattering matrix, were also studied \cite{Kaz}, 
obtaining the same renormalization group equations 
derived for a single node coupled to several semi-infinite 1D wires \cite{Lal}. 
The transport through one-dimensional TLL 
coupled together at a single point has been also studied \cite{Kom}. 
$Y$-junctions of superconducting Josephson junctions were as well analyzed: for suitable values of the control parameters 
an attractive finite coupling fixed point is found \cite{Giul}, displaying an emerging 
two-level quantum system with enhanced coherence \cite{Cirillo}. Star graphs 
were studied also in connection with bosonic models: properties 
of an ideal gas of bosons on a star graph were investigated 
in \cite{Bur01,Bru04} and the possible experimental realization with ultracold 
bosons was discussed in \cite{Bru04}. The dynamics of one-dimensional Bose 
liquids in $Y$-junctions and the reflection at the center of the star was 
studied, discussing the emergence of a repulsive fixed point \cite{Tok}. 
Finally, we mention that the study of different theories on a graph 
and, particularly, on a star graph
is a very active field of research: for example, for the
Laplacian operator (also called quantum graphs) \cite{Ro,ES,KS,Kuc,CR},
 for the Dirac operator \cite{BT,BH}, 
for classical field theories and soliton theories
\cite{SMSSN,CH,ACFN} and for quantum field theories \cite{Bel}.

The plan of the paper is the following: 
in section \ref{sec:xxs}, we introduce the XX model 
on a star graph, and we perform the JW transformation needed to obtain 
a fermionic Hamiltonian. The usefulness to add auxiliary sites is motivated, 
and the obtained Kondo Hamiltonian derived and discussed. In section \ref{sec:ff}, we show 
that it is possible to find an XY model 
such that after the JW transformation one obtains a quadratic fermionic 
Hamiltonian directly diagonalizable. Finally, our conclusions are presented 
in section \ref{sec:conc}.

\section{The XX model on a star graph\label{sec:xxs}}

In this section, we want to obtain fermionic Hamiltonians 
from quantum spin models on a star graph by using a JW transformation.
In particular, we point out the importance to add an 
auxiliary site to obtain a fermionic Hamiltonian: we show that to 
solve this model is equivalent to solve a generalized Kondo model. 
The treatment is explicitly done for the 
XX model to emphasize our construction in the simplest case, although 
the procedure can be used to study other models on the star graph. 

\subsection{The model and the Jordan-Wigner transformation}
\label{model} 

\begin{figure}[htb]
\begin{center}
\includegraphics[width=.45\textwidth]{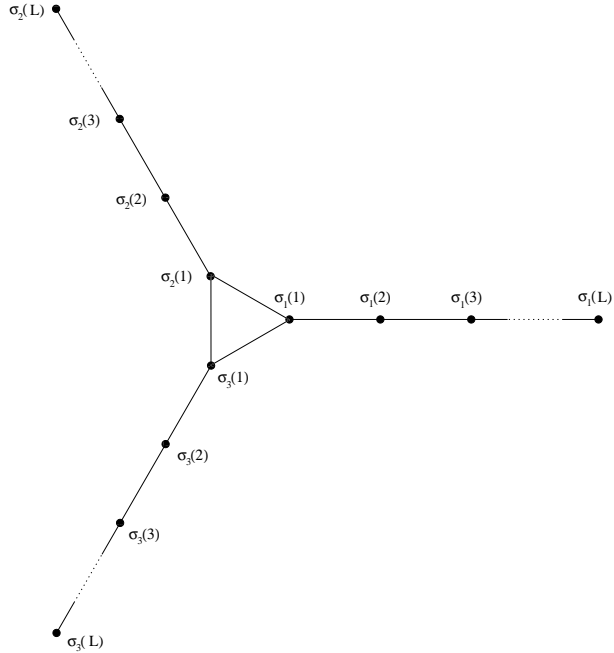}
\caption{The three-leg star graph and the labeling of the vertices.}\label{fig:star}
\end{center}
\end{figure}

We introduce in this section the XX model on a three-leg star graph. 
The graph we consider is illustrated in Figure \ref{fig:star} and it is made 
of three chains of length $L$, each one having vertices labeled 
by $1, \cdots, L$; the sites $1$ of each of the three chains are connected 
between them. In each vertex of the graph (having $3L$ vertices) are defined 
the Pauli matrices $\sigma^x=\left(\begin{array}{c c}
          0&1\\
	  1&0
         \end{array}\right)$, 
$\sigma^y=\left(\begin{array}{c c}
          0&-i\\
	  i&0
         \end{array}\right)$,
$\sigma^z=\left(\begin{array}{c c}
          1&0\\
	  0&-1
         \end{array}\right)$. As usual we use the notation 
$\sigma^\pm=\frac{1}{2}(\sigma^x\pm i \sigma^y)$.

The XX model is described by the following quantum Hamiltonian 
acting on the Hilbert space $(\CC^2)^{\otimes 3L}$:
\bea
\label{eq:H2}
\widetilde{H}^{XX}_3=\sum_{j=1}^{L-1}\sum_{\alpha=1}^3
\sigma^+_\alpha(j)\sigma^-_\alpha(j+1)+
\rho\sum_{\alpha=1}^3 \sigma^+_\alpha(1)\sigma^-_{\alpha+1}(1) + h.c.\:,
\eea
where $\sigma^\pm_\alpha(j)$ stands for the matrix $\sigma^\pm$ 
acting on the $\alpha^{th}$ chain (with $\alpha=1,2,3$) 
and on the $j^{th}$ site from the vertex 
(the labeling of the sites is plotted in Fig. \ref{fig:star}). 
In equation (\ref{eq:H2}) we have used the convention 
$\sigma^\pm_4(1):=\sigma_1^\pm(1)$. 
The parameter $\rho$ (in general complex) entering in the definition of the Hamiltonian 
(\ref{eq:H2}) is a free parameter allowing one to modify the coupling constant at 
the center of the star graph. In particular, for $\rho=0$,
one retrieves three independent XX models on segments with free (open) 
boundaries.

At this point, we arrive at the main ingredient of our construction. 
To perform a JW transformation, we introduce, instead of $\widetilde{H}^{XX}_3$, 
a slightly different Hamiltonian $H^{XX}_3$ acting on the Hilbert space
$\CC^2 \otimes (\CC^2)^{\otimes 3L}$ and defined by the following rules:
$H^{XX}_3$ acts as $\widetilde{H}^{XX}_3$ on 
the last $3L$ $\CC^2$-spaces and trivially on the first $\CC^2$-space.  
The added space (in comparison with the Hilbert space of  $\widetilde{H}^{XX}_3$) 
is denoted $0$ and is called {\em auxiliary space}.
We can write the link between both Hamiltonians as follows:
\begin{equation}
 H^{XX}_3= Id(0) \otimes \widetilde{H}^{XX}_3\;,
\end{equation}
where $Id(0)$ is the 2 by 2 identity matrix acting on the auxiliary space.
Notice that $H^{XX}_3$ has exactly the same spectrum 
as $\widetilde{H}^{XX}_3$ but with a degeneracy multiplied by $2$.
Although the addition of this auxiliary site is trivial 
for the quantum spin model, we will see that it allows one to perform the JW  
transformation to get a fermionic model
(see also the discussion at the end of this section to motivate 
why this auxiliary space seems necessary). The use of 
auxiliary sites to perform a JW transformation 
has been recently used in multidimensional spin systems \cite{Ga,Ci}.

The JW transformation we use is defined, for $j=1,2,\dots,L$, by
\bea
\label{eq:JW}
c_1(j)=\eta^x\left(\prod_{k=1}^{j-1}\sigma_1^z(k)\right)\sigma_1^-(j)~,~~
c_2(j)=\eta^y\left(\prod_{k=1}^{j-1}\sigma_2^z(k)\right)\sigma_2^-(j)~,~~
c_3(j)=\eta^z\left(\prod_{k=1}^{j-1}\sigma_3^z(k)\right)\sigma_3^-(j)\;,
\eea 
where we introduced the following operators:
\bea\label{eq:kf}
\eta^x=\sigma^x(0)\prod_{k=1}^{L}\sigma_2^z(k)\sigma_3^z(k)~,\quad
\eta^y=\sigma^y(0)\prod_{k=1}^{L}\sigma_1^z(k)\sigma_3^z(k)~,\quad
\eta^z=\sigma^z(0)\prod_{k=1}^{L}\sigma_1^z(k)\sigma_2^z(k)\;.
\eea
The last two factors in the r.h.s. of each of the equations (\ref{eq:JW}) 
are the usual JW transformations \cite{JW} and 
give the anti-commutation between terms 
in the same leg. The anti-commutation between different legs is 
provided by the first factor, i.e., by the operators $\eta^a$, with 
$a=x,y,z$, defined by equations (\ref{eq:kf}). The choice (\ref{eq:kf}) 
for the operators $\eta^a$ is due to the need to satisfy the three following 
requests: {\em i)} the operators $c_\alpha(j)$ have to be fermionic; 
{\em ii)} the operator $\eta^a$ has to be $a$-th component of a 
spin operator; {\em iii)} 
the operators $c_\alpha(j)$ and the operators $\eta^a$ have to commute.

Defining as usual $c_\alpha(j)^\dag$ as the conjugate transpose of $c_\alpha(j)$ 
(for $\alpha=1,2,3$ and $j=1,2,\dots,L$), one can indeed show that 
$c_\alpha(j)$ and $c_\alpha(j)^\dag$ are fermionic operators 
[property {\em i})] and that they 
satisfy for $\alpha,\beta=1,2,3$ and $j,k=1,2,\dots,L$ 
the following anti-commutation relations:
\beq
\{c_\alpha(j)\;,\;c_\beta(k)\}=0~,\qquad\{c_\alpha(j)^\dag\;,\;c_\beta(k)^\dag\}=0~,\qquad
\{c_\alpha(j)\;,\;c_\beta(k)^\dag\}=\delta_{\alpha,\beta}\delta_{jk}\;
\eeq
(where $\{.\,,.\}$ stands for the anti-commutator).

Furthermore the operators $\eta^a$ 
share the same relations of the the Pauli matrices [property {\em ii})], since 
they satisfy, for $a,b=x,y,z$, 
\beq\label{eq:ceta}
{\eta^a}^\dag=\eta^a\quad,\quad\{\eta^a,\eta^b\}=2\delta_{ab}\quad\text{and}\quad \eta^x\eta^y=i\eta^z\;.
\eeq
An important point is that $\eta^x$ commutes 
with $\prod_{k=1}^{j-1}\sigma_1^z(k)\ \sigma_1^-(j)$
but anti-commutes with $\prod_{k=1}^{j-1}\sigma_2^z(k)\ \sigma_2^-(j)$. 

Finally, we observe for $a=x,y,z$, $\beta=1,2,3$ and $j=1,2,\dots,L$, 
the following relations hold:
\beq\label{eq:cec}
[\eta^a\;,\;c_\beta(j)]=0\quad\text{and}\quad[\eta^a\;,\;c_\beta(j)^\dag]=0\;
\eeq
according to the requested property {\em iii}).

The factor $\eta^a$ in equations (\ref{eq:JW}) 
may be viewed as a Klein factor, which has been used 
extensively in literature: 
it allows one to define correctly the bosonization \cite{Hal} 
(see also \cite{Gog98,DS,Giam})
and it has been used in different contexts, including 
the 2-channel Kondo model \cite{DFZ}, 
quantum wire junctions described 
by coupled TLL \cite{NFLL,COA} or
the free quantum field theory on a star graph \cite{BMS}.

We conclude this subsection by emphasizing 
that the introduction of the auxiliary site and 
the JW transformation (\ref{eq:JW}) do
not depend on the explicit form of the Hamiltonian. 
Therefore, the construction proposed here may be applied to other models as 
the anisotropic XY model with a transverse magnetic field.

\subsection{The Kondo model}

By using the result of Section \ref{model}, it is possible 
to construct a model equivalent to $H^{XX}_3$ expressed in terms of fermions.
Indeed, by using relations (\ref{eq:JW}), we can express the Hamiltonian $H^{XX}_3$ in terms of
the operators $c_\alpha(j)$, $c_\alpha(j)^\dag$ and $\eta^a$ as follows
\beq\label{eq:Hkondo}
H^{XX}_3=-\sum_{j=1}^{L-1}\sum_{\alpha=1}^3  {c_\alpha(j)}^\dag {c_\alpha(j+1)}
+i\rho\left(
\eta^z \ c_1(1)^\dag c_2(1)
+\eta^x\ c_2(1)^\dag c_3(1)
+\eta^y\ c_3(1)^\dag c_1(1) 
\right)
+h.c.
\eeq
To write more compactly the Hamiltonian (\ref{eq:Hkondo}) 
we introduce $\{S^x,S^y,S^z\}$, the $su(2)$ generators in the 3-dimensional 
representation, as 
\beq
S^x=\left(\begin{array}{c c c}
          0&0&0\\
	  0&0&-i\\
	  0&i&0
         \end{array}\right)~,\qquad
S^y=\left(\begin{array}{c c c}
          0&0&i\\
	  0&0&0\\
	  -i&0&0
         \end{array}\right)~,\qquad
S^z=\left(\begin{array}{c c c}
          0&-i&0\\
	  i&0&0\\
	  0&0&0
         \end{array}\right)\;.
\eeq
Then, for $\rho \in \RR$, the Hamiltonian (\ref{eq:Hkondo}) becomes
\beq\label{eq:Hkondoc}
H^{XX}_3=-\sum_{j=1}^{L-1}\sum_{\alpha=1}^3 \left(
 {c_\alpha(j)}^\dag {c_\alpha(j+1)}+{c_\alpha(j+1)}^\dag {c_\alpha(j)}\right)
-\rho\sum_{a=x,y,z}\sum_{\alpha,\beta=1,2,3} \eta^a\ c_\alpha(1)^\dag (S^a)_{\alpha\beta} c_\beta(1)\;.
\eeq
Finally, by introducing the vectorial notation 
\beq
\boldsymbol{c(j)}^\dag=({c_1(j)}^\dag,{c_2(j)}^\dag,{c_3(j)}^\dag)~,\quad
\boldsymbol{\eta}=\left(
          \eta^x, \eta^y, \eta^z
         \right)~,\quad
\boldsymbol{S}=\left(\begin{array}{c}
          S^x\\
	  S^y\\
	  S^z
         \end{array}\right)\;,
\eeq
the Hamiltonian (\ref{eq:Hkondoc}) 
may be rewritten in a more compact way as 
\beq\label{eq:Hkondoco}
H^{XX}_3=-\sum_{j=1}^{L-1}\left( \boldsymbol{c(j)}^\dag\boldsymbol{c(j+1)}
+\boldsymbol{c(j+1)}^\dag\boldsymbol{c(j)}\right)
-\rho\ \boldsymbol{\eta}\ .\ \boldsymbol{c(1)}^\dag \boldsymbol{S} \boldsymbol{c(1)}  \;.
\eeq
The expression (\ref{eq:Hkondoc}) is valid for three legs 
and it allows us to interpret 
the Hamiltonian $H^{XX}_3$ as the Hamiltonian of 
free fermions coupled with a magnetic impurity.
More precisely, it is a $su(2)$ Kondo model
with free fermions in the spin $1$ representation and 
a magnetic impurity in the fundamental representation.

The historical Kondo model \cite{Kon} - 
studied using, for example, perturbation theory \cite{AHY}, 
numerical renormalization group \cite{Wil} or exact methods \cite{Wieg,And} - 
corresponds to spin $1/2$ free fermions coupled with a spin $1/2$ impurity.
Different generalizations have been introduced and studied: 
spin $S$ impurities \cite{FW},
the $su(N)$ version, 
so-called the Coqblin-Schrieffer model \cite{CS}, 
the multi-channel Kondo models \cite{NB} or the multi-channel 
$su(N)$ fermions in the fundamental representation with a spin $S$ impurity \cite{AZ,JAZ}.

The most relevant results for our case are given in \cite{FG,SK}. 
These papers showed that the dynamics of the spin sector of 
the single channel Kondo model coupling spin $j$ fermions 
with a spin $S$ impurity
is similar to the ones of the $k(j)=2j(j+1)(2j+1)/3$ channel Kondo model
coupling spin $1/2$ fermions with a spin $S$ impurity. 
Then, using the results for the multi-channel Kondo model 
obtained previously by the Bethe 
ansatz method \cite{AD,TW}
and by the conformal field theory \cite{AL}, 
the universal exponents for the thermodynamic 
quantities (e.g. the susceptibility or the specific heat) can be obtained. 
A general discussion of the mapping between different 
multichannel exchange models with spin $j$ fermions, impurity spin $S$ 
and channel number $N_f$ is presented in \cite{FZ96}.
 
These results applied to our case leads to the fact 
that our model (\ref{eq:Hkondoc}),
in the continuum limit,
is equivalent to the $4$-channel Kondo model coupling 
spin $1/2$ fermions with a spin $1/2$ impurity. 
Then, since the number of channels is larger than twice the impurity spin,
we deduce that our system shows a non-Fermi liquid behavior \cite{NB} and that
the impurity contributes, for example, to the susceptibility and the specific,
respectively, as following:
\begin{equation}
 \chi_{imp}\propto T^{-1/3}\qquad\text{and\qquad} C_{imp}\propto T^{2/3}\;.
\end{equation}

We conclude this section by further 
commenting on the JW transformation we used, to 
more clearly show that the standard JW transformation does not generally give 
for a star graph a local and quadratic fermionic Hamiltonian and that 
the JW transformation (\ref{eq:JW})-(\ref{eq:kf}) based on the introduction 
of the space of an auxiliary, fictitious site is functional to have 
the desired commutation relations between the fermionic operators 
and the spin $\boldsymbol{\eta}$. We start by observing 
that one may think of using the JW transformation 
(\ref{eq:jw2}) 
instead of (\ref{eq:JW}), replacing $\eta^a$ by $\sigma^a(0)$. 
However, in this case, although we get a Hamiltonian similar to 
(\ref{eq:Hkondo}),
the fermionic operators obtained do not satisfy the commutation relations 
(\ref{eq:cec}) with the $\eta$'s. Therefore, we cannot 
anymore to directly interpret the model as a Kondo model. 
Furthermore one could also think of different JW transformations 
without adding an auxiliary space. However, such 
transformations have their drawbacks: 
for example, if one performs the transformations (\ref{eq:JW}) without
the first factor (the $\eta$'s), one obtains a quadratic Hamiltonian, 
but quadratic in hardcore bosons, not fermions. Alternatively, 
one could consider a JW transformations following a ``spiral'' according 
\begin{equation}
 \mathfrak{c}_{3(j-1)+\alpha}=\left(\prod_{k=1}^{j-1}\sigma_1^z(k)\sigma_2^z(k)\sigma_3^z(k)\right) 
\left(\prod_{\beta=1}^{\alpha-1}\sigma_\beta^z(j)\right)
\sigma_\alpha^-(j)\quad\text{for }j=1,2,\dots,L~;~~\alpha=1,2,3\;:
\end{equation}
the operators $\mathfrak{c}_j$ are fermionic, however the Hamiltonian finally obtained 
is not quadratic in these operators.

\section{Free fermions on a star graph and associated spin chains\label{sec:ff}}

In this section we investigate if it is possible 
to find a XY model on a star graph which, after a 
JW transformation gives only
a quadratic fermionic Hamiltonian.
In comparison with the previous section, we allow ourselves to modify the interaction 
between the spins near the vertex.

\subsection{Link between Hamiltonians\label{sec:link}}

Since it would be very cumbersome to explore 
all the interactions between spins at the vertex 
to find the ones providing a quadratic fermionic Hamiltonian, 
we proceed in the following way: we start from a quadratic 
fermionic Hamiltonian on the star graph and we perform
a JW transformation, obtaining a quantum spin model on the star graph. 
In this section, we restrict ourselves to the three-leg star graph 
(one may extend easily the obtained results).

We start from the following quadratic fermionic Hamiltonian 
on a three-leg star graph 
\beq\label{eq:H1f}
 \widetilde{H}_3^{QF}=\sum_{j=1}^{L-1} \sum_{\alpha=1}^3\left(
 d_\alpha(j)^\dag d_\alpha(j+1)-\gamma d_\alpha(j) d_\alpha(j+1)\right)
+i\sum_{\alpha=1}^3\left( a_{\alpha}
 d_\alpha(1)^\dag d_{\alpha+1}(1)
+b_{\alpha} d_\alpha(1) d_{\alpha+1}(1)\right)+h.c.\:,
\eeq
where $\gamma$, $a_{\alpha}$ and $b_{\alpha}$ are coupling constants,
$d_\alpha(j)$ and $d_\alpha(j)^\dag$ are fermionic operators and 
we have used the conventions
$d_4(1):=d_1(1)\,,\ d_4^\dag(1):=d_1^\dag(1)$.
As in the previous section, instead of $\widetilde{H}_3^{QF}$ we consider 
the Hamiltonian 
${H}_3^{QF}=Id(0) \otimes \widetilde{H}^{QF}_3$ which trivially acts on a supplementary $\CC^2$-space, 
the auxiliary space denoted by $0$. 
Then, we perform the following JW transformation:
\bea
\label{eq:jw2}
d_1(j)=\sigma^x(0)\prod_{k=1}^{j-1}\sigma_1^z(k)\ \sigma_1^-(j)~,~
d_2(j)=\sigma^y(0)\prod_{k=1}^{j-1}\sigma_2^z(k)\ \sigma_2^-(j)~,~
d_3(j)=\sigma^z(0)\prod_{k=1}^{j-1}\sigma_3^z(k)\ \sigma_3^-(j)\;.
\eea 
with $j=1,2,\dots,L$. A straightforward computation shows that 
the R.H.S. of (\ref{eq:jw2})
are fermionic operators (notice the differences with the previous Jordan-Wigner 
transformations (\ref{eq:JW})).

By using the transformations (\ref{eq:jw2}), 
the quadratic fermionic Hamiltonian becomes the following quantum spin chain
\bea\label{eq:H1}
{H}_3^{QF}=-\sum_{x=1}^{L-1}\sum_{\alpha=1}^3
\left(\sigma^+_\alpha(x)\sigma^-_\alpha(x+1)+\gamma \sigma^-_\alpha(x)\sigma^-_\alpha(x+1)\right)-H_V+h.c.
\eea
where
\bea\label{eq:HVC}
H_V&=&\sigma^z(0)
\left(a_{1}\ \sigma^+_1(1)\sigma^-_{2}(1)
+b_{1}
\ \sigma^-_1(1)\sigma^-_{2}(1)\right)
+\sigma^x(0)
\left(a_{2}\ \sigma^+_2(1)\sigma^-_{3}(1)
+b_{2}
\ \sigma^-_2(1)\sigma^-_{3}(1)\right)\\
&&+\sigma^y(0)
\left(a_{3}\ \sigma^+_3(1)\sigma^-_{1}(1)
+b_{3}
\ \sigma^-_{3}(1)\sigma^-_1(1)\right)
\nonumber
\eea

This shows that it is possible to obtain a XY model 
on a three-leg star graph 
which is equivalent to a quadratic fermionic Hamiltonians. Notice, however, 
that the interactions between spins at the center of the star graph 
are not of the XY type and involve three spins. 

\subsection{Solution for $a_{\alpha}=a$ and $b_{\alpha}=\gamma=0$}

The Hamiltonian $\widetilde{H}_3^{QF}$ given in relation (\ref{eq:H1f}) is a quadratic fermionic Hamiltonian.
Therefore, it can be diagonalized by usual procedures \cite{LSM,LM}. 
Evidently, this result provides also the spectrum of the Hamiltonian 
(\ref{eq:H1}) since it is the same spectrum 
with all the degeneracies multiplied by two.

To present a specific example, we consider in the following 
the case $a_{\alpha}=a\in\RR$, $b_{\alpha}=0$ and $\gamma=0$ 
(for $\alpha=1,2,3$). 
We can rewrite the Hamiltonian (\ref{eq:H1f}) as 
\begin{equation}
\widetilde{H}^{QF}_3=\sum_{i,j=1}^{3L} d_i^{\dag} A_{ij} d_j\:,
\end{equation}
where we have changed the numeration $d_\alpha(j)\rightarrow d_{(\alpha-1)L+j}$
and the entries of the matrix $A$ are $0$ everywhere except
\bea
&&A_{j,j+1}=1=A_{j+1,j}\quad\text{where}~~ j=1,\dots,L-1,L+1,\dots,2L-1,2L+1,\dots,3L-1\\
&&A_{1,L+1}=A_{L+1,2L+1}=A_{2L+1,1}=ia\quad\text{and}\quad 
A_{1,2L+1}=A_{L+1,1}=A_{2L+1,L+1}=-ia\;.
\eea 
To diagonalize $\widetilde{H}^{QF}_3$ one has to 
diagonalize the $3L$ times $3L$ matrix $A$: it is possible to 
show that the eigenvalues of $A$ are the roots of 
the following three polynomials
\beq\label{eq:be}
U_L(\lambda/2)\pm a \sqrt{3}\  U_{L-1}(\lambda/2)=0 ~~\text{or} ~~ U_L(\lambda/2) =0~
\eeq
where $U_L(x)$ is the Chebyshev polynomials of the second kind of degree $L$. 
Let us remark that, for $a=0$, we get three times the same equation which is expected since the 
system becomes three identical decoupled systems.

The Hamiltonian then becomes
\begin{equation}
\widetilde{H}^{QF}_3=\sum_{k=1}^{3L}|\lambda_k|\left(\xi_k^\dag\xi_k-\frac{1}{2}\right)\:,
\end{equation}
where $\xi_k$ are fermionic operators and $\{\lambda_k\}$ is the set of the $3L$ solutions of (\ref{eq:be}).
Therefore, the spectrum of $\widetilde{H}^{QF}_3$ is the following set of $2^{3L}$ elements
\beq
\left\{\frac{1}{2}\sum_{k=1}^{3L}\eps_k |\lambda_k|  \text{ such that } \eps_k=\pm  \right\}
\eeq
and the eigenvalue of the ground state is $-\frac{1}{2}\sum_{k=1}^{3L} |\lambda_k|$.

\begin{figure}[htb]
\begin{center}
\includegraphics[width=.45\textwidth]{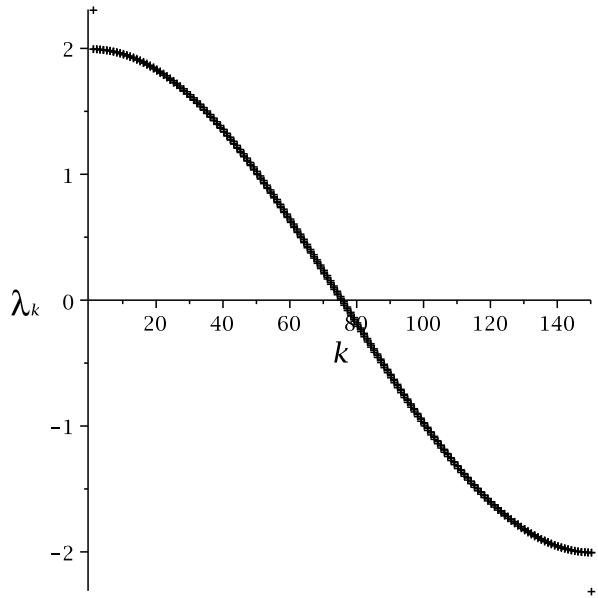}
\caption{The roots of the polynomials (\ref{eq:be}) for $L=150$ and $a=1$.}\label{fi:spec}
\end{center}
\end{figure}

Finally, the problem is solved when the $3L$ solutions of (\ref{eq:be}) 
are given. 
It is easy to find them numerically or even analytically. Indeed, for the last equation, its roots are
$2cos(k\pi/(L+1))$ with $k=1,2,\dots,L$: 
we present them in Figure \ref{fi:spec} 
for $a=1$ and $L=150$ as a dispersion relation.
We remark that the three sets of solutions are very similar 
(in Figure \ref{fi:spec}, these three sets of roots are superimposed). 
The main difference relies on the presence of one 
isolated point for each of the first two equations. 
As shown in subsection \ref{sec:link}, this result provides the spectrum for
the quantum spin model (\ref{eq:H1}).

\section{Conclusions\label{sec:conc}}

In this paper we studied the XX model for quantum spin model on a 
three-leg star graph: we showed that by introducing an auxiliary space 
and performing a Jordan-Wigner 
transformation, the model is equivalent to a generalized Kondo Hamiltonian 
in which the free fermions, in the spin 1 representation of $su(2)$, 
are coupled with a magnetic impurity. Using previous results, we deduce that
it is also equivalent to a 4-channel Kondo model with spin 1/2 fermions coupled 
with spin 1/2 impurity and conclude that it shows a non-Fermi liquid behavior.
We also showed that it is possible to find a XY model 
such that - after the Jordan-Wigner transformation - one obtains a 
quadratic fermionic Hamiltonian directly diagonalizable. 

We observe that we may think of different generalizations 
of our method. Indeed, our method based on the 
Jordan-Wigner transformation (\ref{eq:JW}) 
could be used for the Hamiltonian obtained replacing 
the XX Hamiltonian (\ref{eq:H2}) 
by the anisotropic XY model with a transverse magnetic field. 
In perspective, one can also think to investigate more complicated graphs  
as the star graph with a number of legs $p>3$ or as 
comb-like graphs: in order to get a Kondo-like Hamiltonian, i.e., an 
Hamiltonian of fermions coupled with magnetic impurities, one should 
identify the correct Klein factors, which in the present case $p=3$ 
are given by equations (\ref{eq:kf}). 
Furthermore, the properties of a quantum Ising model in a transverse 
magnetic field on a graph ${\cal G}$ may be related to the partition 
function of the classical Ising model in a corresponding 
higher-dimensional graph. Namely, the partition function of the classical 
Ising model on a graph made up of $n$ copies of the ``base'' graph ${\cal 
G}$ with couplings between corresponding sites of the adjacent copies can 
be written as the trace of the transfer matrix $V$ to the power $n$ where 
$V$ is written as exponentials of terms proportional to $\sigma^z \, 
\sigma^z$ and $\sigma^x$. 
By organizing the factors in the exponentials, we may recognize the exponential of 
a quantum Ising model.
For example, for a square (resp. cubic) lattice, 
the ``base'' graph ${\cal G}$ is a line (resp. square): e.g., for the 
square, $V$ can be written in terms of the quantum Ising model in a 
transverse magnetic field on the segment \cite{LSM2}. Therefore getting 
results for Kondo Hamiltonians of type (\ref{eq:Hkondoco}) obtained from 
quantum Ising models on star-like graphs may be relevant to study the 
classical Ising model in non trivial geometries.

The mapping presented in this paper 
between the XX quantum spin model on the star graph and the Kondo model 
illustrates that the introduction of a non trivial topology, even locally, 
can provide new interesting physical phenomena in
comparison to models on the line or on the circle. At the same time, 
our results show that one may also think to use the XX model 
on a star graph to realize (or simulate) a Kondo model: to this respect 
we mention that a similar Hamiltonian, describing 
Majorana fermions, can be realized in a superconductor, coupled 
to conduction electrons \cite{BC}.

\vspace{0.5cm}

{\em Acknowledgments:} We would like to thank V. Caudrelier, D. Giuliano, 
M. Fabrizio, P. Sodano and P.B. Wiegmann for very useful discussions. 
Useful correspondences with B. Beri, P. Lecheminant, D.C. Mattis and A.M. Tsvelik are also gratefully 
acknowledged. 

\vspace{0.5cm}

{\em Note Added:} After this paper was submitted, several 
very interesting papers on $Y$-systems appeared on the arXiv. 
In \cite{T} the problem of an Ising model in a transverse field has 
been studied on the star graph: in the continuum 
limit, close to the quantum phase transition point 
and for coupling $\rho <<1$, the 
effective Lagrangian was worked out and the model shown to be equivalent 
to the overscreened two-channel Kondo model \cite{T}. As previously mentioned, 
the approach presented in our paper can be used for the 
general case of an anisotropic XY in a transverse field: it would then 
very interesting to study the Kondo problem in such more general model, 
determining in particular how the low-energy physics varies across 
the parameter space (i.e, varying the anisotropy and the magnetic field). 
A discussion of the coupling of Majorana fermions 
to external leads was presented in \cite{Ber12}: 
the Klein factors of bosonization 
appear as extra Majoranas hybridizing with the physical ones and a 
$SO(M)$ Kondo problem was shown to arise \cite{Ber12}. In 
\cite{altland12} it was studied a setup with 
nanowires in proximity to a common mesoscopic 
superconducting island, showing that a weak finite charging energy 
of the center island may considerably affect 
the low-energy behavior of the system. Finally, in \cite{zazunov12} 
a general Majorana junction was considered and 
the conditions for even-odd parity effects in the tunnel conductance 
for various junction topologies were examined.


\begin{thebibliography}{10}

\bibitem{Cha11}S. Chakrabarty and Z. Nussinov,
\textsl{Modulation and correlation lengths in systems with competing interactions,}
Phys. Rev. B \textbf{84} (2011) 144402.

\bibitem{Bur00}R. Burioni, D.Cassi, I. Meccoli, M. Rasetti, S. Regina, P. Sodano, and A. Vezzani, 
\textsl{Bose-Einstein condensation in inhomogeneous Josephson arrays,}
Europhys. Lett. \textbf{52} (2000) 251.

\bibitem{Sod06}P. Sodano, A. Trombettoni, P. Silvestrini, R. Russo,
and B. Ruggiero, 
\textsl{Inhomogeneous superconductivity in comb-shaped Josephson 
junction networks,}
New J. Phys. \textbf{8} (2006) 327. 

\bibitem{Mat06}D.C. Mattis,  
\textsl{The theory of the magnetism made simple} 
(World Scientific, 2006).

\bibitem{Lac11} 
\textsl{Introduction to frustrated magnetism: Materials, experiments, theory,} 
eds. C. Lacroix, P. Mendels, and F. Mila (Heidelberg, Springer, 2011).

\bibitem{Igl}F. Igl\'oi, I. Peschel, and L. Turban, 
\textsl{Inhomogeneous systems with unusual critical behaviour,}
Adv. Phys. \textbf{42} (1993) 683.

\bibitem{Kon}J. Kondo,
\textsl{Resistance minimum in dilute magnetic alloys,}
Prog. Theor. Phys. \textbf{32} (1964) 37.

\bibitem{Hew93}A.C. Hewson, 
\textsl{The Kondo problem to heavy fermions} 
(Cambridge, Cambridge University Press, 1993).

\bibitem{Kou01}L. Kouwenhoven and L. Glazman, 
\textsl{Revival of the Kondo effect,} 
Phys. World \textbf{14} (2001) 33.

\bibitem{Ali96}A.P. Alivisatos, 
\textsl{Semiconductor clusters, nanocrystals, and quantum dots,} 
Science \textbf{271} (1996) 933.

\bibitem{Kou98}L. Kouwenhoven and C.M. Marcus, 
\textsl{Quantum dots,}
Phys. World \textbf{11} (1998) 35.

\bibitem{Laf08}N. Laflorencie, E.S. Sorensen, and I. Affleck,
\textsl{Kondo effect in spin chains,}
J. Stat. Mech. P02007 (2008).

\bibitem{Aff08}I. Affleck, 
\textsl{Quantum impurity problems in condensed matter physics,}
\texttt{arXiv:0809.3474}

\bibitem{Bay12}A. Bayat, S. Bose, P. Sodano, and H. Johannesson, 
\textsl{Entanglement probe of two-impurity Kondo physics in a spin chain,}
Phys. Rev. Lett. \textbf{109} (2012) 066403.

\bibitem{Bay10}A. Bayat, P. Sodano, and S. Bose, 
\textsl{Negativity as the entanglement measure to probe the Kondo regime in the spin-chain Kondo model,}
Phys. Rev. B \textbf{81} (2010) 064429.

\bibitem{Sod10}P. Sodano, A. Bayat, and S. Bose, 
\textsl{Kondo cloud mediated long range entanglement after local quench in a spin chain,}
Phys. Rev. B \textbf{81} (2010) 100412(R).

\bibitem{CardyBook}J.L. Cardy, 
\textsl{Scaling and renormalization in statistical physics} 
(Cambridge, Cambridge University Press, 1996).

\bibitem{Cardy}J.L. Cardy, 
\textsl{Critical behaviour at an edge,} 
J. Phys. A \textbf{16} (1983) 3617.

\bibitem{JW}P. Jordan and E. Wigner, 
\textsl{\"{U}ber das Paulische \"{A}quivalenzverbot,}
Z. Physik \textbf{47} (1928) 631.

\bibitem{LSM2}T.D. Schultz, D.C. Mattis, and E.H. Lieb,
\textsl{Two-dimensional Ising model as a soluble problem of many fermions,}
Rev. Mod. Phys. \textbf{36} (1964) 856. 

\bibitem{Muss}G. Mussardo,
\textsl{Statistical field theory: An introduction to exactly solved models in statistical physics} 
(Oxford, Oxford University Press, 2010).

\bibitem{DM}D.C. Mattis,
\textsl{Soluble Ising model in 2+1/N dimensions and 
XY model in 1+1/N dimensions,}
Phys. Rev. B \textbf{20} (1979) 349.

\bibitem{Ga}V. Galitski, 
\textsl{Fermionization transform for certain higher-dimensional 
quantum spin models,}
Phys. Rev. B \textbf{82} (2010) 060411.

\bibitem{Ci}F. Verstraete and J.I. Cirac,
\textsl{Mapping local Hamiltonians of fermions to local Hamiltonians of spins,} 
J. Stat. Mech. P09012 (2005).

\bibitem{Lal}S. Lal, S. Rao, and D. Sen, 
\textsl{Junction of several weakly interacting quantum wires: A renormalization group study,}
Phys. Rev. B \textbf{66} (2002) 165327.

\bibitem{COA}C. Chamon, M. Oshikawa, and I. Affleck,
\textsl{Junctions of three quantum wires and the dissipative Hofstadter model,} 
Phys. Rev. Lett. \textbf{91} (2003) 206403; 
\textsl{Junctions of three quantum wires,}
J. Stat. Mech. P02008 (2006).

\bibitem{Giul}D. Giuliano and P. Sodano, 
\textsl{$Y$-junction of superconducting Josephson chains,} 
New J. Phys. \textbf{10} (2008) 093023; 
\textsl{Frustration of decoherence in $Y$-shaped superconducting Josephson networks,} 
Nucl. Phys. B \textbf{811} (2009) 395.

\bibitem{Kaz}K. Kazymyrenko and B. Doucot, 
\textsl{Regular networks of Luttinger liquids,}
Phys. Rev. B \textbf{71} (2005) 075110.

\bibitem{Kom}A. Komnik and R. Egger, 
\textsl{Nonequilibrium transport for crossed Luttinger liquids,} 
Phys. Rev. Lett. \textbf{80} (1998) 2881.

\bibitem{Cirillo}A. Cirillo, M. Mancini, D. Giuliano, and P. Sodano,
\textsl{Enhanced coherence of a quantum doublet coupled to Tomonaga-Luttinger liquid leads,}
Nucl. Phys. B \textbf{852} (2011) 235.

\bibitem{Bur01}R. Burioni, D. Cassi, M. Rasetti, P. Sodano, and A. Vezzani,
\textsl{Bose-Einstein condensation on inhomogeneous complex networks,}
J. Phys. B \textbf{34} (2001) 4697.

\bibitem{Bru04}I. Brunelli, G. Giusiano, F.P. Mancini, 
P. Sodano, and A. Trombettoni,
\textsl{Topology induced spatial Bose-Einstein condensation 
for bosons on star-shaped optical networks,}
J. Phys. B \textbf{37} (2004) S275.

\bibitem{Tok}A. Tokuno, M. Oshikawa, and E. Demler, 
\textsl{Dynamics of one-dimensional Bose liquids: Andreev-like reflection at $Y$ junctions and the absence of the Aharonov-Bohm effect,}
Phys. Rev. Lett. \textbf{100} (2008) 140402.

\bibitem{Ro}J.P. Roth,
\textsl{Le spectre du Laplacian sur un graphe,} in 
\textsl{Th{\'e}orie du potentiel - Lecture notes in mathematics,} 
eds. A.D. Dold and B. Beckmann, pp. 521-539, in french (Springer-Verlag 1983).

\bibitem{ES}P. Exner and P. \v{S}eba,
\textsl{Free quantum motion on a branching graph,}
Rep. Math. Phys. \textbf{28} (1989) 7.

\bibitem{KS}V. Kostrykin and R. Schrader,
\textsl{Kirchoff's rule for quantum wires,} 
J. Phys. A \textbf{32} (1999) 595.

\bibitem{Kuc}P. Kuchment, 
\textsl{Quantum graphs I. Some basic structures,} 
Waves in Random Media \textbf{14} (2004) S107. 

\bibitem{CR}V. Caudrelier and E. Ragoucy,
\textsl{Direct computation of scattering matrices for general quantum graphs,}
Nucl. Phys. B \textbf{828} (2010) 515. 

\bibitem{BT}W. Bulla and T. Trenkler, 
\textsl{The free Dirac operator on compact and noncompact graphs,} 
J. Math. Phys. \textbf{31} (1990) 1157.

\bibitem{BH}J. Bolte and J.Harrison,
\textsl{Spectral statistics for the Dirac operator on graphs,}
J. Phys. A \textbf{36} (2003) 2747.

\bibitem{SMSSN}Z. Sobirov, D. Matrasulov, K. Sabirov, 
S. Sawada, and K. Nakamura,
\textsl{Soliton solutions of nonlinear Schr\"odinger equation on simple networks,}
Phys. Rev. E \textbf{81} (2010) 066602.

\bibitem{CH}R.C. Cascaval and C.T. Hunter,
\textsl{Linear and nonlinear Schr\"odinger equations on simple networks,}
Libertas Mathematica \textbf{30} (2010) 85.

\bibitem{ACFN}R. Adami, C. Cacciapuoti, D. Finco, and D. Noja,
\textsl{Fast solitons on star graphs,}
Rev. Mat. Phys. \textbf{23} (2011) 409.

\bibitem{Bel}B. Bellazzini, M. Burrello, M. Mintchev, and P. Sorba, 
\textsl{Quantum field theory on star graphs,}
Proc. Symp. Pure Math. \textbf{77} (2008) 639.
 
\bibitem{Hal}F.D.M. Haldane, 
\textsl{Coupling between charge and spin degrees of freedom in the one-dimensional Fermi gas with backscattering,}
J. Phys. C \textbf{12} (1979) 4791.

\bibitem{Gog98}A.O. Gogolin, A.A. Nersesyan, and A.M. Tsvelik, 
\textsl{Bosonization and strongly correlated systems}  
(Cambridge, Cambridge University Press, 1998).

\bibitem{DS}J. von Delft and H. Schoeller,
\textsl{Bosonization for beginners --- Refermionization for experts,}
Annalen Phys. \textbf{7} (1998) 225.

\bibitem{Giam}T. Giamarchi, 
\textsl{Quantum physics in one dimension} 
(Oxford, Oxford University Press, 2004).

\bibitem{DFZ}J. von Delft, M. Fabrizio, and G. Zar\'and,  
\textsl{Finite-size bosonization of 2-channel Kondo model: A bridge between numerical renormalization group and conformal field theory,} 
Phys. Rev. Lett. \textbf{81} (1998) 196.

\bibitem{NFLL}C. Nayak, M. Fisher, H. Lin, and A. Ludwig, 
\textsl{Resonant multilead point-contact tunneling,}
Phys. Rev. B \textbf{59} (1999) 15694.

\bibitem{BMS}B. Bellazzini, M. Mintchev, and P. Sorba,
\textsl{Bosonization and scale invariance on quantum wires,}
J. Phys. A \textbf{40} (2007) 2485.

\bibitem{AHY}P.W. Anderson, D.R. Hamann, and G. Yuval,
\textsl{Exact results in the Kondo problem. II. Scaling theory, qualitatively Correct solution, and some new results on one-dimensional classical statistical models,}
Phys. Rev. B \textbf{1} (1970) 4464. 

\bibitem{Wil}K.G. Wilson,
\textsl{The renormalization group: Critical phenomena and the Kondo problem,}
Rev. Mod. Phys. \textbf{47} (1975) 773. 

\bibitem{Wieg}P.B. Wiegmann,
\textsl{Exact solution of s-d exchange model at T=0,}
JETP Lett. \textbf{31} (1980) 364.

\bibitem{And}N. Andrei, 
\textsl{Diagonalization of the Kondo Hamiltonian,}
Phys. Rev. Lett. \textbf{45} (1980) 379.

\bibitem{FW}V.A. Fateev and P.B. Wiegmann,
\textsl{The exact solution of the s-d exchange model with arbitrary spin S,}
Phys. Lett. A \textbf{81} (1981) 179.

\bibitem{CS}B. Coqblin and J.R. Schrieffer,
\textsl{Exchange interaction in alloys with Cerium impurities,}
Phys. Rev. \textbf{185} (1969) 847.  

\bibitem{NB}P. Nozi\`{e}res and A. Blandin,
\textsl{Kondo effect in real metals,}
J. Phys. France \textbf{41} (1980) 193. 

\bibitem{AZ}N. Andrei and P. Zinn-Justin,
\textsl{The generalized multi-channel Kondo model: Thermodynamics and fusion equations,}
Nucl. Phys. B \textbf{528} (1998) 648.

\bibitem{JAZ}A. Jerez, N. Andrei, and G. Zar\'and,
\textsl{Solution of the multichannel Coqblin-Schrieffer impurity model and application to multi-level systems,} 
Phys. Rev. B \textbf{58} (1998) 3814.

\bibitem{FG} M. Fabrizio and A.O. Gogolin,
\textsl{Toulouse limit for the overscreened four-channel Kondo problem,} 
Phys. Rev. B \textbf{50} (1994) 17732.
    
\bibitem{SK}A.M. Sengupta and Y.B. Kim,
\textsl{Overscreened single channel Kondo problem,}
Phys. Rev. B \textbf{54} (1996) 14918.

\bibitem{AD}N. Andrei and C. Destri,
\textsl{Solution of the multichannel Kondo problem,}
Phys. Rev. Lett. \textbf{52} (1984) 364.

\bibitem{TW}A.M. Tsvelik and P.B. Wiegmann,
\textsl{Solution of the n-channel Kondo problem (scaling and integrability)},
Z. Phys. B \textbf{54} (1984) 201.

\bibitem{AL}I. Affleck and A.W.W. Ludwig,
\textsl{The Kondo effect, conformal field theory and fusion rules,}
Nucl. Phys. B \textbf{352} (1991) 849;
\textsl{Critical theory of overscreened Kondo fixed points,}
Nucl. Phys. B \textbf{360} (1991) 641;
\textsl{Exact conformal-field-theory results on the multichannel 
Kondo effect: Single-fermion Green’s function, self-energy, and resistivity,}
Phys. Rev. B \textbf{48} (1993) 7297.

\bibitem{FZ96}M. Fabrizio and G. Zar\'and,
\textsl{Mapping between multichannel exchange models,}
Phys. Rev. B \textbf{54} (1996) 10008.

\bibitem{LSM}E. Lieb, D.C Mattis, and T.D. Schultz, 
\textsl{Two soluble models of an antiferromagnetic chain,} 
Ann. Phys. \textbf{16} (1961) 407.

\bibitem{LM}E. Lieb and D.C. Mattis, 
\textsl{Mathematical Physics in One Dimension} 
(New York and London, Academic Press, 1966).

\bibitem{BC}B. B\'eri and N.R. Cooper,
\textsl{Topological Kondo effect with Majorana fermions,}
 Phys. Rev. Lett. \textbf{109} (2012) 156803.

\bibitem{T}A.M. Tsvelik,
\textsl{Realization of 2-channel Kondo effect 
in a junction of three quantum Ising chains,}
\texttt{arXiv:1211.3481}

\bibitem{Ber12}B. B\'eri, 
\textsl{Majorana-Klein hybridization in topological superconductor junctions,}
\texttt{arXiv:1212.4465}

\bibitem{altland12}A. Altland and R. Egger,
\textsl{Multi-terminal Coulomb-Majorana junction,}
\texttt{arXiv:1212.6224}


\bibitem{zazunov12}A. Zazunov, P. Sodano, and R. Egger, 
\textsl{Even-odd parity effects in Majorana junctions,}
\texttt{arXiv:1301.6882}    
     

\end{thebibliography}
\end{document}